\documentclass[12pt]{article}
\usepackage{epsfig,graphicx}

\title{The quark-hadron thermodynamics in magnetic field}
\author{  V.D.Orlovsky and Yu.A.Simonov \\ Institute of Theoretical and Experimental
Physics\\ 117218, Moscow, B.Cheremushkinskaya 25, Russia}

\newcommand{\be}{\begin{equation}}
\newcommand{\ee}{\end{equation}}

\def\la{\mathrel{\mathpalette\fun <}}

\def\fun#1#2{\lower3.6pt\vbox{\baselineskip0pt\lineskip.9pt
\ialign{$\mathsurround=0pt#1\hfil ##\hfil$\crcr#2\crcr\sim\crcr}}}

\newcommand{{\SD}}{\rm SD}

\newcommand{\vep}{\mbox{\boldmath${\rm p}$}}

\newcommand{\vez}{\mbox{\boldmath${\rm z}$}}

\newcommand{\veB}{\mbox{\boldmath${\rm B}$}}

\newcommand{{\Mc}}{\mathcal{M}}

\newcommand{\lan}{\langle}
\newcommand{\ran}{\rangle}

\begin{document}

\maketitle
\begin{abstract}
Nonperturbative treatment of quark-hadron transition  at nonzero temperature
$T$ and chemical potential $\mu$ in the framework of Field Correlator Method is
generalized to the case of nonzero magnetic field B.  A compact  form of the
quark pressure for arbitrary $B,\mu, T$ is derived. As a result the transition
temperature is found as a function of B  and  $\mu$, which depends on only
parameters: vacuum gluonic condensate $G_2$ and the field correlator
$D_1^E(x)$, which defines the Polyakov loops and it is known both analytically
and on the  lattice. A moderate (25\%) decrease of $T_c(\mu=0)$ for
$eB$ changing  from zero to 1 GeV$^2$ is found.  A sequence of transition
curves in the $(\mu, T)$ plane is obtained for $B$ in the same interval,
monotonically decreasing in scale for growing $B$.
\end{abstract}
\section{Introduction}

Strong magnetic fields (m.f.) are now a subject of numerous studies
\cite{1*,2*,3*,4*,5*,6*,7*,8*}, since they can be present in different physical
systems. Namely, in cosmology m.f. of the order of $10^{18}$ Gauss or higher
can occur during strong and electroweak phase transition  \cite{1*,2*}. In
noncentral heavy ion collisions one can expect m.f. $O(10^{18}- 10^{21})$ Gauss
\cite{3*,4*, 5*, 6*}, while in  some classes of neutron stars m.f. can reach
the magnitude of $10^{14}$ Gauss, or even more in the cental regions \cite{7*}.
All this makes it necessary to study the effects of strong m.f. in all possible
physical situations and using different methods, for a recent review see
\cite{{8}*}.

One of most interesting aspects of strong m.f. is its influence on the QCD
hadron-quark phase transition, which can occur both in astrophysics (neutron
stars) and heavy ion experiments. On the theoretical side many model QCD
 calculations have predicted the increase of the critical temperature $T_c$
with growing B \cite{9*,10*,11*,12*,13*,14*,15*,16*,17*,18*,19*,20*,21*,22*,23*,24*,25*,26*,27*,28*,29*,30*,31*,32*,33*,34*,35*},
and only  few obtained an opposite result \cite{36*,37*}, see \cite{37a,37**}
for reviews and additional references. Recently the lattice data of \cite{38*}
with physical pion mass and extrapolated to continuum have demonstrated the
decreasing critical temperature as a function of  $B$. This phenomenon was
called the inverse magnetic catalysis and the further study of the m.f.
dependence of the  quark condensate and of its magnetic susceptibility was done
in \cite{38a} and  \cite{38**} respectively. It is our purpose in this paper to
exploit the formalism of Field Correlators (FC) developed earlier for the QCD
phase transition at zero m.f. \cite{1,2,3,4,5,6,7} to  study the same problem in
the case of arbitrary m.f.

The advantage of the FC method  is that it is based  only on the  fundamental
QCD input: gluonic condensate $\lan G^2\ran$, string tension  $\sigma$,
$\alpha_s(q)$ and current quark masses. In contrast to \cite{36*}, where the
same basic principle \cite{2} was used, but pions were elementary in CPTh, it
treats all hadrons, including pions, as $q\bar q$ or $3q$ systems, which allows
to consider high m.f. with $eB\gg m^2_\pi, $ $\sigma$. As a result the critical
temperature $T_c(B)$  decreases  with the growing  $B$ as in lattice data  of
\cite{38*}.

Recently the FC method was successfully applied to the study of phase
transition in neutron stars \cite{49*,50*} without  m.f. and in the case of
strange quarks and strange matter in \cite{51*}. It is interesting to
investigate the role of m.f. in these transitions and our results below may be
a reasonable starting point for this analysis.

The paper is organized as follows. In  section 2 the general FC formalism as
applied to the hadron-quark transition is given, and in section 3 the
contribution of magnetic field is explicitly taken into account. In section 4
the quark and hadron thermodynamic potentials are estimated at large m.f. and
the corresponding transition  temperature $T_c(B)$ is found. In section 5 the
case of nonzero chemical potential is treated and  in section 6 a discussion of
results and prospectives is presented.

\section{General formalism}

We shall follow the ideas of \cite{1,2,3,4,5,6,7} (see \cite{8} for a review)
and consider the low-temperature hadron phase as the hadron gas in the
confining background vacuum field, and the total free energy can be represented
as
\be
F=\varepsilon_{vac} V_3 + F_h,\label{1}
\ee
where
\be
\varepsilon_{vac}
= \frac{\beta(\alpha_s)}{16 \alpha_s} \lan G^a_{\mu\nu} G^a_{\mu\nu} \ran +
\sum_q m_q \lan \bar q q\ran,\label{2}\ee
and $F_h$ is the hadron free energy,
which in absence of magnetic field  and treating hadrons as elementary can be
written as \cite{9,10} $(\beta =1/T)$
\be
- F_{h/V_3} = \sum_i  P_h^{(i)} =
\sum_i \frac{g_i T}{2\pi^2} \int^\infty_0 dp p^2 \eta \ln \left(1 + \eta
e^{-\beta E_i}\right),\label{3}
\ee
where $\eta =-1$ or $+1$ for bosons or
fermions respectively and in the relativistic case $E_i = \sqrt{\vep^2 +
m^2_i}$, while $g_i $ is the spin-isospin multiplicity of hadron $i$.
 Taking integral in (\ref{3}), one can write $P_h^{(i)}$ as
 \be P_h^{(i)} = \frac{g_i T^4}{2\pi^2} \sum^\infty_{n=1} (- \eta)^{n+1}
 \frac{(\beta m_i)^2}{n^2} K_2 (n\beta m_i),\label{5}\ee where $K_2$ is the Mc
 Donald function.
 As an example of another starting point we present below the derivation of the
 quark pressure from the statistical sum in the form of the generating function
 with the proper time integration  \cite{3,5,6,7}.
 \be \frac{1}{T} F_q =\frac12 \ln \det (m^2_q - \hat D^2) = -\frac12 tr
 \int^\infty_0 \xi (s) \frac{ds}{s} e^{-s m^2_q + s\hat D^2}\label{6}\ee

 The latter expression can be written as a path integral with the background
 field containing both electromagnetic $A^{(e)}_\mu (x)$ and color potential
 $A_\mu (x) $, \cite{8,11,12,13}
 \be \frac{1}{T} F_q (A,A^{(e)} ) = - \frac12 tr \int^\infty_0 \xi (s)
 \frac{ds}{s} d^4 x\overline{(Dz)}^w_{xx} e^{-K-sm_q^2} \lan W_\sigma
 (C_n)\ran, \label{7}\ee
where $K=\frac14 \int^s_0 \left( \frac{dz_\mu}{d\tau} \right)^2 d \tau$,

\be W_\sigma (C_n) = P_F P_A \exp (ig \int_{C_n} A_\mu dz_\mu + ie \int_{C_n}
A_\mu^{(e)} dz_\mu) \exp \int^s_0 (g\sigma_{\mu\nu} F_{\mu\nu} +
e\sigma_{\mu\nu} F_{\mu\nu}^{(e)})  d\tau,\label{8}\ee and  \be
\overline{(Dz)}^w_{xy} = \prod^n_{m=1} \frac{d^4\Delta
z_k(m)}{(4\pi\varepsilon)^2}\sum_{n=0,\pm 1,\pm 2} (-)^n \frac{d^4p}{(2\pi)^4}
e^{ip_\mu(\sum\Delta z_\mu(m) - (x-y) - n \beta \delta_{\mu 4})}.\label{9}\ee
It was shown in \cite{6,7}, that $P^{(i)}_q \equiv P_q$ can be written as \be
P_q = 2 N_c\int^\infty_0  \frac{ds}{s} e^{-m^2_q s} \sum^\infty_{n=1}
(-)^{n+1}[S^{(n)} (s) + S^{(-n)} (s)]\label{10}\ee and \be S^{(n)} (s) = \int
(\overline{Dz} )^w_{on} e^{-K} \frac{1}{N_c} tr  W_\sigma (C_n)\label{11}\ee
and in the case, when only one-particle contribution is retained, \be S^{(n)}
(s) = \frac{1}{16 \pi^2s^2} e^{-\frac{n^2\beta^2}{4s}- J^E_n},\label{12}\ee and
$J^E_n$ defines the Polyakov loop configuration expressed via the field
correlator $D_1(x) $ \cite{6,7,14} \be J^E_n = \frac{n\beta}{2} \int^{n\beta}_0
d\nu \left( 1- \frac{\nu}{n\beta}\right) \int^\infty_0 \xi d\xi D^E_1
(\sqrt{\nu^2+ \xi^2}),\label{13}\ee where $D^E_1(x)$ is the colorelectric
correlator, which stays nonzero above the deconfinement temperature.

The insertion of (\ref{12}) into (\ref{10}) yields  $$\frac{1}{T^4} P_q =
\frac{N_c n_f}{4\pi^2} \sum^\infty_{n=1} \frac{(-)^{n+1}}{n^4}  \int^\infty_0
\frac{ds}{s^3} e^{-m^2_qs - \frac{n^2\beta^2}{4s} - J_n^E}=$$ \be= \frac{4N_c
n_f}{\pi^2} \sum^\infty_{n=1}\frac{(-)^{n+1}}{n^4}  \varphi^{(n)}_q L^n,
~~\varphi^{(n)}_q = \frac{n^2 m^2_q }{2T^2} K_2 \left(\frac{m_qn}{T}\right),
\label{14}\ee where $L^n=e^{-J^E_n}$

At this point it is convenient to  give one more representation of $P_q$,
namely, using \cite{13} one can extract in $z_4(\tau)$ the fluctuating part
$\tilde z_4 (\tau)$ \be z_4 (\tau) = \bar z_4 (\tau) + z_4(\tau), ~~\bar z_4
(\tau) = 2\omega \tau =t_E,~~ s=\frac{T_4}{2\omega}, ~~ T_4=n\beta\label{15}\ee
\be P_q = 2 N_c n_f \int^\infty_0 \frac{d \omega}{\omega}
\sqrt{\frac{\omega}{2\pi}}\sum^\infty_{n=1}\frac{(-)^{n+1}}{\sqrt{n\beta}} \int
D^3ze^{-K(w)-J^E_n} \label{16}\ee \be K(\omega) = \int^{n\beta} _0 dt_E \left(
\frac{\omega}{2}+ \frac{ m^2}{2\omega} +
\frac{\omega}{2}\left(\frac{d\vez}{dt_E}\right)^2\right), ~~ m_q\equiv
m.\label{17}\ee

In this way we obtain the form, equivalent to (\ref{14})\be P_q=
\frac{N_cn_f}{\pi^2\beta^2} \sum^\infty_{n=1}\frac{(-)^{n+1}}{n^2}\int^\infty_0
\omega d\omega e^{-\left( \frac{m^2}{2\omega} + \frac{\omega}{2}\right) n\beta
-J^E_n}.\label{18}\ee Neglecting $J^E_n$ and for $m_q=0$ one obtains the
standard result \be P_q = \frac{4N_c n_f T^4}{\pi^2}
\sum^\infty_{n=1}\frac{(-)^{n+1}}{n^4}= \frac{7N_c n_f \pi^2}{180}
T^4,\label{19}\ee where we have used \be \int^\infty _0 \omega d\omega
e^{-\left( \frac{m^2}{2\omega} + \frac{\omega}{2}\right) n\beta }= 2m^2 K_2
(mn\beta).\label{20}\ee For the following it will be useful to keep in
(\ref{16}) the $d^3p$ integration, contained in $D^3z$, which yields \be P_q =
\frac{n_f N_c}{\sqrt{\pi}} \int \frac{d^3 p}{(2\pi)^3}\sum^\infty_{n=1}
{(-)^{n+1}}\sqrt{\frac{2}{n\beta}} \int^\infty_0
\frac{d\omega}{\sqrt{\omega}}e^{-\left( \frac{m^2+\vep^2}{2\omega} +
\frac{\omega}{2}\right) n\beta  }.\label{21}\ee

One can see, that (\ref{21})  coincides with (\ref{5}), when $g_i = 4 N_c n_f$.
Finally, one obtains from (\ref{5}) or (\ref{19}) the pressure for gluons \be
P_g = (N_c^2-1) \frac{2T^4}{\pi^2} \sum^\infty_{n=1} \frac{\lan\Omega^n\ran +
\lan \Omega^{*n}\ran}{2n^4},\label{22}\ee where $\Omega^n=\exp (ig
\int^{n\beta}_0 A_4 dz_4$), and $\Omega$ is the adjoint Polyakov loop.

\section{Quark and hadron thermodynamics in magnetic field}

We discuss here the one-particle thermodynamics in constant homogeneous m.f.
$\veB$ along $z$ axis, in  which case one should replace $E_i$ in (\ref{3}) by
the well-known expression  \cite{10}, which in the relativistic case has the
form \be E^\sigma_{n_\bot} (B) = \sqrt{ p^2_z + (2n_\bot +1 - \bar \sigma)|e_q|
B+ m_q^2},~~~~ \bar \sigma\equiv \frac{e_q}{|e_q|} \sigma_z, \sigma_z =\pm
1.\label{23}\ee We also take into account, that the phase space of  an isolated
quark in m.f. is changed as follows \cite{9}\be \frac{V_3d^3p}{(2\pi)^3} \to
\frac{dp_z}{2\pi} \frac{|e_qB|}{2\pi} V_3,  \label{24}\ee and hence (\ref{3})
can be rewritten as $ \bar P_q (B) = \sum_q P_q (B),$

\be
P_q (B) = \sum_{n_\bot, \sigma} 2 N_c   T \frac{|e_qB|}{2\pi}\frac12
(\chi(\mu)+\chi(-\mu)),\label{25}
\ee
where
\be
\chi(\mu)\equiv \int \frac{dp_z}{2\pi} \ln
\left(1+\exp \left( \frac{\bar \mu-E^\sigma_{n_\bot}(B)}{T}\right)
\right).\label{25*}
\ee

We have introduced  in(\ref{25}) the chemical potential  $\mu$ with the
averaged Polyakov loop factor $\bar L=\exp(-\bar J/T)$, (see \cite{8} for a
corresponding treatment without m.f.)  \be \bar \mu=\mu - \bar J, ~~ \bar
L_\mu= \exp \left(\frac{\mu-\bar J}{T}\right).\label{26}\ee  Eq.(\ref{25}) can
be integrated over $dp_z$  with the result \be P_q (B) = \frac{N_c  |e_q
B|T}{\pi^2} \sum_{n_\bot, \sigma} \sum^\infty_{n=1} \frac{(-)^{n+1}}{n} \frac12
 (\bar L^n_\mu+\bar L_{-\mu}^n) \varepsilon^\sigma_{n_\bot} K_1 \left( \frac{n
\varepsilon^\sigma_{n_\bot}}{T}\right),\label{27}\ee where \be
\varepsilon^\sigma_{n_\bot} = \sqrt{ |e_qB| ( 2n_\bot +1-\bar{\sigma}) +
m^2_q}.\label{28}\ee

The same result  for $\mu=0$ can be  obtained, extending (\ref{21}) to the case
of nonzero $B$, using (\ref{24}) and replacing the exponent in (\ref{21}) as
\be \left( \frac{m^2_q+\vep^2}{2\omega} + \frac{\omega}{2}\right)  n\beta \to
\left( \frac{m^2_q + p^2_z + ( 2 n_\bot +1 -\bar{\sigma}) |e_qB| }{2\omega} +
\frac{\omega}{2}\right) n\beta,\label{29}\ee \be P_q = 2  N_c \frac{|e_q
B|}{(2\pi)^2} \sum^\infty_{n=1} \frac{(-)^{n+1}}{n\beta} \int^\infty_0 d\omega
e^{-\left(\frac{(\varepsilon\sigma)^2}{2\omega} + \frac{\omega}{2}\right)
n\beta},\label{30}\ee and using the equation \be  \int^\infty_0 d\omega
e^{-\left( \frac{\lambda^2}{2\omega} +\frac{\omega}{2}\right)  \tau} = 2\lambda
K_1 (\lambda\tau),\label{31}\ee we come to the Eq. (\ref{27}).

We turn now to thermodynamics of hadrons in m.f. The difficulty here is that
hadrons are not elementary objects, unlike quarks, and we cannot use for them
the energy expressions like (\ref{23}). Hadrons in m.f. were studied
analytically in \cite{15,16,17,18} and on the lattice in \cite{19,20,38a}. We
can use for them an expression of the type of Eq. (\ref{25}) or Eq. (\ref{27}),
however we should write it in a more general way for the charged hadrons
\be
P_H^{(i)} (B) = g_i \frac{|e_HB|T}{2\pi} \int \frac{dP_z}{2\pi} \frac12
(\chi(\mu_i) + \chi(-\mu_i)), \label{32}
\ee
where
\be
\chi(\mu_i)\equiv \sum_{n_\bot, s_i} \ln
\left(1+\exp \left( \frac{\mu_i - E_{N_\bot}^{s_i} (B)}{T} \right)
\right), \label{32*}
\ee
and we take into account, that the total hadron energy
$E^{s_i}_{N_\bot} (B)$ depends on the set of 2d oscillator numbers $N_\bot =
\{n_\bot (1), n_\bot (2), ... n_\bot (\nu)\}$ for each of $\nu$ constituents
and on the set $s_i = \{ \sigma_1, ... \sigma_\nu\}$ of spin projections of all
constituents.  For very large m.f. $eB\gg \sigma (\sigma$ is the string
tension) one can approximate $E^{s_i}_{N_\bot} (B)$ as an average of the sum of
constituents (2 for mesons and 3 for baryons), \be E_{N_\bot}^{s_i} (B) =
\left\lan \sum^\nu_{k=1} \sqrt{ m^2_q(k) + |e_k| B(2n_\bot (k) +1 -
\frac{e_k\bar\sigma_k}{|e_k|}) + p^2_z(k)}\right\ran_{p_z}\label{33}\ee where
the average is taken with the functions $ \chi (p_z (1), ... p_z (\nu))$,
satisfying $\sum^\nu_{k=1} p_z (k) = P_z$, and taking into account confining
dynamics along $z$ axis (see explicit expressions in the Appendix).  In the
approximation used in \cite{15,16,17,18}, when confinement is quadratic, the
functions $\chi$ are the  oscillator eigenfunctions. For neutral hadrons one
should use instead of (\ref{32}) the form (\ref{5}), where m.f. acts on the
multiplicity $g_i$ and the mass $m_i$, which can strongly depend on m.f., as it
is  in the case of $\rho_0$ and $\pi^0$ mesons, see \cite{16,17}.

At this point it is useful to compare the systematics of hadrons without m.f.
with that of strong m.f. In the first case one classifies   a hadron, using
e.g. spin $S$, partly $P$ isospin $I$, orbital momentum $L$ and radial quantum
number $n_r$, or else total angular momentum $J$. For strong m.f. both spin $S$
(or $J)$ and isospin are not conserved and one has e.g. instead of  2 states
$\rho^0(S_z=0), \pi^0$ linear combinations  $\lan u+, \bar u-|\equiv \lan + -
|_u, ~~\lan -+|_u, ~~ \lan +-|_d, ~~ \lan -+|_d$ and similarly   $\rho^0
(S_z=+1)$ splits into 2 states: $\lan ++ |_u, ~~ \lan ++|_d$.

A similar situation occurs in baryons: neutron,  $S_z= -\frac12, (ddu)$ splits
into $(--+), (-+-), (+--).$

A specific role is here played by the so-called ``zero states'':  those are
 states for which all constituents have   factors in (\ref{33}) equal to  zero:
 \be 2n_\bot (k) + 1 - \frac{e_k}{|e_k|}  \bar{\sigma}_k =0, ~~
 k=1,2,...\nu.\label{34}\ee

 Masses of zero states decrease fast with m.f. and for $eB\approx \sigma$ can be
 $(30\div 40)\%$ lower  than in absence of m.f. \cite{16,17}. Therefore the role of these
 states in the  forming of $P_h (B)$ exponentially grows, while  energies
 $E^{s_i}_{N_i} (B)$ of all other states according to (\ref{33}) grow
 proportionally to $\sqrt{eB}$. Thus in $\rho^0 (S_z=0), \pi^0$ only  the
 states $\lan +-|_u, \lan -+|_d$ are zero states, while for  the neutron with $S_z
 =- \frac12$ the only zero state is $(--+)$. In this way the  most part of all
 excited hadron states have energies growing with m.f.  and  their contribution
 is strongly suppressed for $eB>\sigma$. However, the same  situation occurs
 for the system of free quarks at large m.f., which can be clearly seen
 comparing (\ref{33}) with energies of free quarks, therefore the  main
 difference occurs for not large m.f., when $eB<\sigma$ and hadron energies
 change less rapidly than  those of free  quarks, and  hence $P_q$ may grow
 faster with $eB$ than $P_h$, which finally   results in the decreasing $T_c
 (B)$, as we show below.

 Indeed, for each quark the zero states constitute one half of $n_\bot=0$
 states, namely the states with $\bar \sigma_k =1$, and the corresponding pressure is proportional to
$ \frac{|e_q| B m_qT}{n} K_1 \left( \frac{m_qn}{T} \right),$ tending  to $
\frac{|e_q|BT^2}{n}$ at large m.f., thus growing  linearly with $B$.

For hadrons in the same limit the charged zero states contribute in (\ref{32})
the amount $\frac{|e_h|BT^2}{n}$, while neutral zero states, like $\pi^0$,
contribute $\frac{T^4}{n^4}$. Therefore for $B\geq T^2$ the growth of quark
pressure  with $B$ is faster then that of hadrons, and one can assert, that for
$B\geq T^2, \sigma$ the inequality holds \be \Delta P_q (B,T) \equiv P_q (B,T)
- P_q (0,T) > \Delta P_h (B,T)= P_h (B,T) - P_h (0,T).\label{35}\ee

 In the   next section we shall show, that (\ref{35}) leads to the decreasing of the
deconfinement temperature $T_c(B)$ with growing $B$ independently of the
character of this transition. In particular, the above arguments were based on
the single-line approximation for quarks \cite{6,7}, when quarks are treated as
independent and vacuum fields create only Polyakov line contributions ($\bar
L^n$ in (\ref{27})). A more accurate treatment, taking into account the $q\bar
q$ interaction due to the $D_1$ correlator (cf Eq. (\ref{13})), shows, that the
$q\bar q$ pairs can form bound states in this interaction \cite{14}, and with
increasing m.f. the binding energy grows, which lowers the $q\bar q$ mass, thus
leading to the growth of the quark pressure. Moreover, the  introduction of
this  ``intermediate state of deconfinement'', existing in the narrow region
near $T_c$, consisting of bound and decaying $q\bar q$ pairs strongly affects
the nature of the deconfining transition, making it softer. In addition, the
colorelectric string tension, which disappears at $T_c$, decreases  gradually
in the same region, lifting in this way the hadron pressure and making the
transition continuous. However this remark does not change qualitatively the
considerations of the present paper and will be treated in detail elsewhere.

\section{The quark-antiquark contribution to the pressure at nonzero m.f.}

As it was discussed above, the nonperturbative $D_1$ contribution to a single
quark is given by Eqs. (\ref{12}), (\ref{13}), where $J^E_n$ can also be
written as

\be L_{ fund} =\exp (-J^E_1)=\exp \left(
-\frac{V_1(\infty)}{2T}\right),\label{36}\ee where $V_1(r)$  is the $q\bar q$
nonperturbative (np) colorelecric interaction generated by the field correlator
$D_1^E (x)$ \cite{14}

\be V_1 (r,T) = \int^\beta_0 (1-\nu T)     d\nu\int^r_0 \xi d\xi D^E_1
(\sqrt{\nu^2+\xi^2}).\label{37}\ee

As it was argued
 in \cite{22}, the asymptotics of $D_1^E(x)
$ is expressed
 via the gluelump mass $M_0\approx 1$ GeV \cite{23} and can be  written  as
\be D_1^{(np)} (x) = \frac{A_1}{|x|} e^{-M |x|} + O(\alpha^2_s),~~
A_1=2C_2\alpha_s \sigma_{adj} M_0,\label{38}\ee which leads to \be V_1(r,T) =
V_1(\infty,T) - \frac{A_1}{M^2_0} K_1(M_0r) M_0r
+O\left(\frac{T}{M_0}\right),\label{39}\ee and \be V_1(\infty, T) =
\frac{A_1}{M_0^2}\left[ 1-\frac{T}{M_0}(1-e^{-M_0/T})\right], ~~
\frac{A_1}{M^2_0} \approx \frac{6\alpha_s (M_0)\sigma_f}{M_0} \approx 0.5 {\rm
GeV}.\label{40}\ee At $r=0,V_1(0,T) =0$.

Above $T_c$ the value of $V_1(\infty, T)$ is decreasing, as seen from
(\ref{40}), (\ref{39}). This is in agreement with lattice data on Polyakov
loops in \cite{24}. Recently, the potential $V_1(r,T)$ was studied on the
lattice in \cite{25}, yielding a behavior similar  for $V_1(\infty, T)$ at
$T=1.2 T_c$.

We now consider the hadron and quark-gluon pressure in the single-line (the
independent particle) approximations with
 the purpose to define the deconfinement temperature as a function of m.f.

One starts with the total pressure in the  confined phase, phase I, which can
be written in the form, generalizing the results of \cite{1,2,3} for the case
of nonzero m.f.

\be P_I =|\varepsilon_{vac}| + \sum_i P^{(i)}_H (B), \label{41}\ee where
$\varepsilon_{vac} = \varepsilon^{(g)}_{vac} + \varepsilon^{(q)}_{vac} $ is
given in (\ref{2}), and we assume, that the gluonic condensate does not depend
on m.f. in the first approximation, while
 the quark condensate $|\lan \bar q q\ran|$
 grows with m.f., as  shown analytically in \cite{26} and on the lattice
\cite{38a,38**}, however  we neglect this contribution  in the first
approximation and discuss its importance  at large $eB$ is the concluding
section.

In the deconfined phase (phase II) the pressure can be written in the form (cf
\cite{1,2,3,6})  \be P_{II} =\frac12 |\varepsilon_{vac}^{(g)}| + \sum_q P_q (B)
+ P_g\label{42}\ee where $P_q(B)$ is given in (\ref{25})-(\ref{26}), and we
assume, that vacuum colormagnetic fields, retained in the deconfined phase
 at $T\approx T_c$, create one-half of vacuum condensate $G_2$ as it happens for $T=0$.

\be \frac12|\varepsilon_{vac}^{(g)} | \cong \frac{(11-\frac23 n_f)}{32} \Delta
G_2, ~~ \Delta G_2 \approx \frac12 G_2.\label{43}\ee

Taking into account the chemical potential $\mu$, one can rewrite (\ref{27}) as
 \be P_q (B) =\sum_{q=u,d,...} \frac{N_c  |e_qB| T}{\pi^2} \sum_{n_\bot,
\sigma=\pm 1}\sum^\infty_{ n=1} \frac{(-)^{n+1}}{n} \cosh \frac{\mu n}{T}
L^n_{fund} \varepsilon^\sigma_{n_\bot} K_1
\left(\frac{n\varepsilon^\sigma_{n_\bot}}{T}\right)\label{44}\ee

Finally, for gluon pressure we are neglecting the influence of m.f., which
appears in higher $O(\alpha_s)$ orders, and write \be P_{gl}(B)\cong
P_{gl}^{(0)} =\frac{2(N^2_c-1)}{\pi^2} \sum^\infty_{n=1} \frac{T^4}{n^4}
L^n_{adj}.\label{45}\ee

As a result we define the deconfinement temperature from the equality \be
P_I(T=T_c)=P_{II} (T=T_c).\label{46}\ee

The contribution of zero levels of light quarks clearly  dominates in
(\ref{44}), when $eB>T^2$, so that keeping for simplicity only the $n=1$,
$\sigma=1$ terms for small $\mu$ , one has

\be \bar P_q(B) \approx P_q^{(0)} (B) = \frac{N_cn_f |\bar e_q B|T^2}{\pi^2}
L_{fund} \cosh\frac{\mu}{T},\label{47}\ee where$ |\bar{e}_q| = \frac{e}{n_f}
\sum^{n_f}_{i=1} \frac{|e_i|}{e}$, ~ $\bar e_q = \frac49 e$
for $n_f=3$.

Neglecting as a first approximation the hadron pressure  and
$\varepsilon^{(q)}_{vac}$in (\ref{41}), one obtains an equation for $T_c$:

\be \frac12|\varepsilon_{vac}^{(g)}|  = P_{gl}^{(0)} +   P^{(0)}_q (B),
\label{48}\ee and finally, neglecting the term $\sum_q m_q |\lan \bar q q\ran
|$, and $P_{gl}^{(0)}$ for large $B>T^2$, one obtains the asymptotic expression

\be T^2_c =\frac{(11-\frac23 n_f) G_2 \pi^2  }{64 N_c n_f | \bar e_q| B
L_{fund} \cosh \frac{\mu}{T}}.\label{49}\ee

 For $\mu=0$, we take $L_{fund} =\exp \left(-\frac{V_1}{2T_c}\right), ~~ V_1 \approx 0.5 $ GeV  $ \cite{14},  n_f =3,~~ \bar e_q
=\frac49 e, ~~ G_2= 0.006$ GeV$^4$   \cite{27}, and we obtain \be T_c (eB= 1 {\rm GeV}^2) \cong 0.125~{\rm GeV}.\label{50}\ee

 For the same parameters and $B=0$ in \cite{27} one gets $T_c(0) \cong 0.165$
 GeV.
 These values are in a
reasonable agreement with the corresponding lattice data in \cite{38**},
$T_c^{lat} (1$ GeV$^2) \approx 0.138 $ GeV, $T_c^{lat} (0) \approx 0.16$ GeV.

 One can now take into account at large $B$ also the contribution of gluons,
 $P_{gl} = \frac{16}{\pi^2}\exp \left(-\frac{9V_1}{8T} \right)$ and $\pi^0$
 mesons, $P_{\pi^0} \approx \frac{\pi^2}{90} T^4$ since the mass of $\pi^0$
 tends to zero for $eB>\sigma$, while that of $\pi^+, \pi^-$ grows as  $\sqrt{eB}$ and does not
 contribute appreciably to the pressure. One can check that solving equation
 \be \frac12|\varepsilon_{vac}^{(g)} | + P_{\pi^0} = P^{(0)}_{gl} +
 P_q^{(0)} (B)\label{51}\ee
 one obtains $T_c(eB =1 $GeV$^2$), which is 2\% larger, than in (\ref{50}).

 At large $eB\gg T^2$  and $\mu=0, ~~ G_2 =0.006$ GeV$^4$,  Eq.(\ref{49})
 yields
 \be T_c^2 = \frac{0.00205~{\rm GeV^2}}{eB} \exp \left(\frac{0.25{\rm
 ~GeV}}{T_c}\right)\label{51a}\ee and one obtains $T_c(eB=6~{\rm GeV^2})\approx
 0.08 $ GeV and a slow decrease for  larger $eB$, $T_c(eB)\sim \frac{1}{\ln (eB)} .$

 To investigate the behavior  of $T_c(B)$  at all values  of $B$ and $\mu=0$ we
  take into account  all Landau levels, as it was done in the Appendix, and
  write resulting expression for $P_q (B)$ in the case $\mu=0$(cf. Eq.
  (\ref{A5})),

 $$ P_q (B) = \frac{N_c e_q BT}{\pi^2} \sum^\infty_{n=1} \frac{(-)^{n+1}}{n}
\bar L^n \left\{ m_q K_1 \left( \frac{nm_q}{T}\right)+\right.$$ \be \left.+
\frac{2T}{n} \frac{e_qB+m_q^2}{e_qB} K_2 \left( \frac{n}{T} \sqrt{e_q B + m^2_q}\right) - \frac{ne_q
B}{12T} K_0\left( \frac{n}{T} \sqrt{m^2_q + e_q B}\right)\right\}\label{52a}\ee

In the limiting case of small quark mass, $m_q\ll \{T, ~ e_qB\}$ one can rewrite
(\ref{52a}) as \be P_q (B) \approx \frac{ N_c e_q BT}{\pi^2} \bar L \left\{ T +
2T K_2 \left( \frac{\sqrt{e_q B}}{T} \right) - \frac{e_qB}{12T} K_0 \left(
\frac{\sqrt{e_qB}}{T}\right)\right\}.\label{53a}\ee

Note, that  Eq. (\ref{53a}) for small $eB$ tends to the limiting
$B$-independent form (\ref{14}), (\ref{19}). We shall use the forms
(\ref{52a}), (\ref{53a}) at all values of $eB$, and hence recalculate
(\ref{48}) with $P_q^{(0)}\equiv \bar P_q(B) = \sum_{q=u,d,s} P_q (B)$, and
$P_q(B)$ from (\ref{52a}), (\ref{53a}).

 As a result from Eq. (\ref{48}) one obtains the curve $T_c(B)$ shown in Fig.~1
 together with the points obtained on the lattice in \cite{38*}.

 \begin{figure}[h]
  \centering
  \includegraphics[width=9cm]{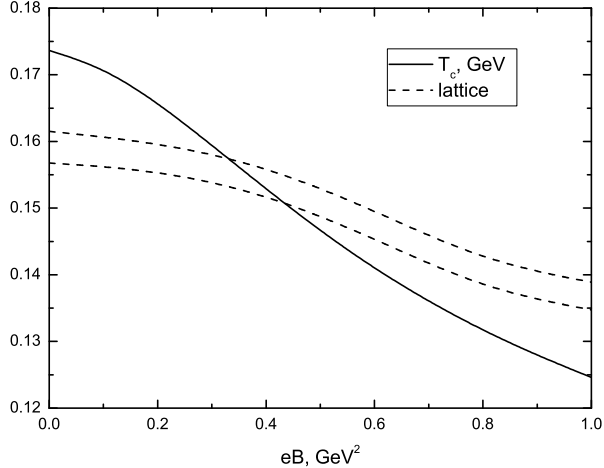}
  \caption{QCD phase diagram in $B-T$ plane for $\mu=0$ as it is given by (\ref{48}) in comparison with lattice data \cite{38*}.}
\end{figure}

 \section{The case of nonzero chemical potential }

 We first consider the case of very large $eB$, when one can retain only the
 lowest Landau levels of quarks.

 For nonzero $\mu$ and $e_q B\gg 4T^2$  one can keep only  zero Landau level and  rewrite (\ref{25}) in the form

 \be P_q (B) = \frac{N_c }{2\pi^2}   e_q B (\phi (\mu) + \phi (-\mu)),
 \label{52}\ee
 where $\phi( \mu)$ is
 \be \phi(\mu) = \int^\infty_0 \frac{p_zdp_z}{1+e\frac{p_z-\bar \mu}{T}}, ~~
 \bar \mu=\mu-\bar J \equiv \mu - \frac{V_1(\infty)}{2}.\label{53}\ee

 At large $\frac{\bar \mu}{T} \gg 1$ one can use the expansion \cite{10}
 \be \phi (\mu) \approx \frac{(\mu-\bar J)^2}{2}+ \frac{\pi^2}{6}
 T^2;\label{54}\ee
 and one  obtains in the lowest approximation (neglecting $\pi^0$ and gluon
 contribution at large  $eB$), which yields ($\frac12 |\varepsilon^{(g)}_{vac}| =
 P_q(B))$ the critical value of the chemical potential $\mu_c$,

 \be \bar \mu^2_c = \frac{1.386 G_2}{\bar e_qB}; ~~ \mu_c = \frac{V_1(\infty)}{2} +
 1.18 \sqrt{\frac{G_2}{eB}}.\label{55}\ee

 For $eB\to \infty $ one has $\mu_c (eB\to \infty) = \frac{V_1(\infty)}{2}
 =0.25$ GeV,
 where we have  assumed, that  $V_1(\infty)$ is independent  of $eB$.

 Near the critical  point the critical curve is easily obtained from
 (\ref{52}), (\ref{54}).

\be  \left( \mu_c - \frac{V_1(\infty)}{2}\right)^2 + \frac{\pi^2}{3} T^2_c =
\frac{1.386 G_2}{eB}.\label{56}\ee

 For small $\mu$ the  lattice data  of \cite{66} reveal, that $V_1(\infty)$   depends
on $eB$ and may become negative  for large $T$ and $B$.

However, for  small $T$ the behavior of $L$ in \cite{66} is compatible with our
assumption, that $V_1(\infty)$  is   weakly dependent on $T$ and being  around
0.5 GeV, which supports our form of the phase transition curve (\ref{56}). Also
the  $\mu$ dependence of the color screening potential $V_1(\infty, \mu)$ was
studied in \cite{25}, and was found to be rather moderate for $(\mu/T)^2 \leq
1$ and $T/T_c = 1.20$ and 1.35. Therefore we can assume, that the behavior
(\ref{56}) is qualitatively correct for large $eB$, $eB >\sigma =0.18$ GeV$^2$,
and it should go over for $B\to 0$  into the form found earlier in \cite{27}.

 Now we turn to the case of arbitrary m.f. As shown in the appendix, one can
 sum up in (\ref{25}) over $n_\bot,
\sigma$ in the following way \be P_q (B) = \frac{N_ce_q B}{2\pi^2} \left\{
\phi(\mu) + \phi(-\mu) +\frac23 \frac{(\lambda (\mu) + \lambda (-\mu))}{e_qB} -
\frac{e_q B}{24} (\tau(\mu)+ \tau(-\mu))\right\},\label{59a}\ee where
$\phi(\mu)$ is given in (\ref{53}), while $\lambda(\mu), \tau(\mu)$ are \be
\lambda(\mu) = \int^\infty_0 \frac{p^4dp}{\sqrt{p^2+ \tilde m^2_q}}
\frac{1}{\exp \left( \frac{\sqrt{ p^2+\tilde m^2_q}-\bar
\mu}{T}\right)+1},\label{60}\ee

 \be
\tau(\mu) = \int^\infty_0 \frac{dp_z}{\sqrt{p^2+ \tilde m^2_q}} \frac{1}{\exp
\left( \frac{\sqrt{ p^2_z+\tilde m^2_q}-\bar \mu}{T}\right)+1}.\label{61}\ee

Here $\tilde \mu^2_q= m^2_q+e_qB$. One can see, that $\lambda(\mu), \tau(\mu)$
decay exponentially for $e_qB\to \infty$, and hence one returns to
Eq.(\ref{52}) in this limit. In the opposite case, when $e_qB\to0$, one
recovers the form (\ref{59a}) with only $\lambda(\mu)+\lambda(-\mu)$ present,
which exactly coincides with one, studied in \cite{27}.

One can now calculate  the transition curve $T_c(\mu, B)$ in the $(T, \mu)$
plane for different values of $eB$, using (\ref{59a}) for $\bar P_q(B) =\sum_q
P_q (B)$ in the equation $\frac12|\varepsilon_{vac}^{(g)}| = P_g^{(0)} +\bar
P_q(B)$. The resulting sequence of
curves  $T_c (\mu, eB), ~~ 0\leq\mu\leq \mu_c$ and $eB =(0,0.2,0.6,1)$ GeV$^2$
is shown in Fig 2. One can see from (\ref{55}), that asymptotically for large
$eB$ the limiting curve is a cut of the straight line $T_c=0, ~~ 0\leq \mu\leq
\frac{V_1(\infty)}{2}.$ Of special importance is the region of small $eB, eB\la
4T^2$, where the asymptotic (large $eB$) $P_q (B,\mu)$ from Eq. (\ref{52})
turns over into the m.f. independent  form of Eq. (\ref{14}).

The details of this transition are given in the Appendix.

 \begin{figure}[h]
  \centering
  \includegraphics[width=9cm]{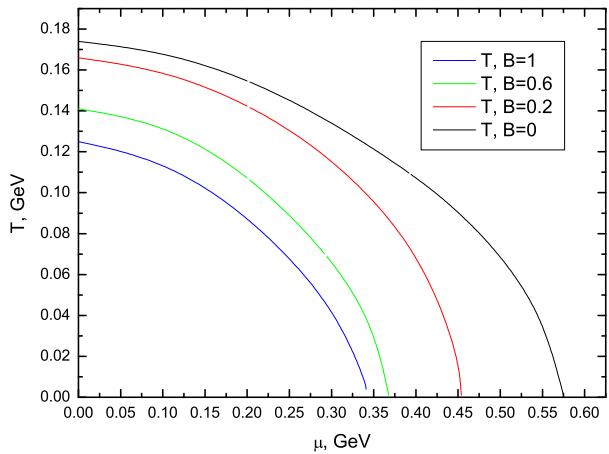}
  \caption{QCD phase diagram in $\mu-T$ plane for different values of $eB$.}
\end{figure}

\section{Discussion of results and conclusions}

We have derived in the paper the compact forms of the quark   pressure for zero
chemical potential and arbitrary $T$ and $eB$ in (\ref{52a}) and for nonzero
chemical potential in (\ref{59a}), which take into account higher Landau
levels.

Using that, we have calculated $T_c(eB, \mu=0)$ in  Eq. (\ref{49}) and Fig.~1
and $T_c(eB, \mu)$ in Fig.~2 in the lowest approximation, neglecting hadron
contributions, except for $\pi^0$, and neglecting possible dependence of  gluon
vacuum energy $\varepsilon^{(g)}_{vac}$ and Polyakov loop $\bar L$ on  m.f.
This approximation, which can be however a crude one, is supported by available
lattice data on the  m.f. dependence of interquark potential $V_1(\infty)$
\cite{25, 66}.

We have shown, that $T_c(eB, \mu=0)$ is a moderately decreasing function of
$eB$, tending to zero asymptotically as $\frac{1}{\ln eB}$. This fact agrees
reasonably well with the realistic lattice data \cite{38*}. Qualitatively the
decreasing pattern was obtained in \cite{36*} and \cite{37*}. However, in both
cases $T_c$ dropped much faster, and  in \cite{36*} $T_c$  passes zero at
$eB\approx 0.7$ GeV$^2$.

A common feature of all approaches, resulting in the decreasing $T_c (eB)$, is
that  they  contain a constant piece of pressure in the  confinement phase,
which is destructed by transition to the quark phase: it is the vacuum
condensate in the present paper and \cite{36*}  and the MIT bag pressure in
\cite{37*}. However the treatment of the hadronic pressure in all available
papers,  assumes that hadrons are elementary and  their masses are  not
modified by m.f. In contrast to that, our approach suggests to use the real
composite hadrons, with masses  strongly dependent on m.f., as was found in
\cite{16}. In the present paper we have used this notion only marginally,
taking $\pi^0$ into account, massless at large $eB$, and neglecting heavy in
this limit $\pi^+, \pi^-$. However, in the next paper we plan to return to this
problem and  to calculate $T_c (eB, \mu)$ with the lowest mass hadrons.

In the  present paper we have used the only parameters of our approach:
$|\varepsilon^{(g)}_{vac}|$ and $V_1 (\infty)$, the latter defining the
Polyakov loop average: $\bar L = \exp \left(-\frac{V_1(\infty)}{2 T}\right)$.
For $|\varepsilon^{(g)}_{vac}|$ we have  used $G_2 \equiv \frac{\alpha_s}{\pi}
\lan F^a_{\mu\nu} F^a_{\mu\nu}\ran = 0.006 $ GeV$^4$, which was found to give
the realistic $T_c (\mu=0, B=0) \approx 165$ MeV, in  reasonable agreement with
most lattice data for $n_f =3$.

The range of values of $G_2$ including $G_2 = 0.006$ GeV$^4$ was studied in
\cite{49*,50*} and found to give reasonable values of stable mass configuration
for a hybrid star configuration. From the purely theoretical point of view, the
values of $G_2$ are not uniqely defined, and the value 0.006  GeV$^4$ is within
the boundaries of  the analysis  in  \cite{73}.

We have neglected the dependence of vacuum parameter $|\varepsilon^{(g)}_{vac}|
$ on $B, \mu$, since   this dependence can occur only in higher orders of
$\alpha_s$ expansion. However we disregarded the quark component $\varepsilon_{vac}^{(q)}
\equiv \sum_q m_q \lan \bar q q\ran$ of the vacuum
energy, $|\varepsilon_{vac}| =
|\varepsilon^{(q)}_{vac}|+|\varepsilon_{vac}^{(g)}|$, assuming the limit $m_q\to 0$. In reality
the contribution of the  strange quark with $m_s (2 {\rm GeV}) \cong 0.1$ GeV
is significant and grows linearly with $eB$, which changes the asymptotics of
$T_c (eB)$ at large $eB$ from $1/\ln(eB)$ to a constant one. Thus the results of
the present paper refer to the case of the $n_f =3$ massless quarks, and we
plan to extend our results to the realistic (2+1) case in the next publication.

Concerning the $\mu$- dependence of $T_c(\mu, B)$ one can see in Fig. 2 a
smooth decreasing behavior for an increasing $B$, with a limiting piece of
straight line $ T_c (\mu, B \to \infty) \to 0, $~~ $\mu_c \to
\frac{V_1(\infty)}{2} $.

We have not analyzed in the paper the character of the quark-hadron transition,
since it needs a careful analysis of the hadronic phase, and possible change of
parameters below $T_c$, which  will be published elsewhere.

The authors  are grateful to N.O.~Agasian, M.A.~Andreichikov, A.M.~Badalian and
B.O.~Kerbikov for  useful suggestions and  discussions.

\vspace{2cm}
 \setcounter{equation}{0}
\renewcommand{\theequation}{A \arabic{equation}}

\hfill {\it  Appendix  }

\centerline{\bf \large The quark thermodynamics in a weak m.f.}

 \vspace{1cm}

\setcounter{equation}{0} \def\theequation{A \arabic{equation}}

 We start with the case of $\mu=0$. Eq.(\ref{27}) can be rewritten as $\bar P_q (B)=\sum_q
 P_q(B),~~ e_q\equiv |e_q|,$
 \be P_q (B)= \frac{N_c e_qB T}{\pi^2} \sum^\infty_{n=1} \frac{(-)^{n+1}}{n}
 \bar L^n \sum_{n_\bot,\sigma} \varepsilon^\sigma_{n_\bot} K_1
 \left(\frac{n\varepsilon^\sigma_{n_\bot}}{T}\right),\label{A1}\ee
 where $\varepsilon^\sigma_{n_\bot}$ is given in (\ref{28}). The sum over
 $n_\bot, \sigma$ can be transformed as follows
 \be
 \sum^\infty_{n_\bot=0} \sum_{\sigma\pm 1} \varepsilon^\sigma_{n_\bot}K_1
 \left(n\varepsilon^\sigma_{n_\bot}{T}\right) = m_q K_1 \left( \frac{nm_q}{T}
 \right) + 2 \sum^\infty_{n_\bot=0} \bar \varepsilon K_1 \left(\frac{n\bar
 \varepsilon}{T}\right).\label{A2}\ee
 Here $\bar \varepsilon = \sqrt{m^2_q + e_q B + 2 e_q B (n_\bot+\frac12)}$.

 At this point one can use the well-known approximation for the summation in the
 limit of weak m.f. (see \S 59 of \cite{10})
 \be \sum^\infty_{k=0} F (k+\frac12) \approx \int^\infty_0 F(x) dx +
 \frac{1}{24} F'(0).\label{A3}\ee
 In the integral in (\ref{A3}) one can use Eq. (\ref{20}) and Eq.(\ref{31}) to write
 $$
 \int^\infty_0 dx \sqrt{m^2_q + e_qB + 2 e_q Bx}~  K_1\left( \frac{n\sqrt{m^2_q +
 e_q B + 2 e_qBx}}{T}\right)=$$
\be\frac{T}{2 ne_q B} \int^\infty_0 \omega d \omega e^{-\frac{n}{T} \left(
\frac{\omega}{2} + \frac{e_qB+ m^2_q}{2\omega}\right)} = \frac{T}{n} \frac{e_qB+m_q^2}{e_qB} K_2 \left(
\frac{n}{T} \sqrt{e_q B + m^2_q}\right).\label{A4}\ee

As a result (\ref{A1}) can be rewritten as
$$ P_q (B) = \frac{N_c e_q BT}{\pi^2} \sum^\infty_{n=1} \frac{(-)^{n+1}}{n}
\bar L^n \left\{ m_q K_1 \left( \frac{nm_q}{T}\right)+\right.$$ \be \left.+
\frac{2T}{n} \frac{e_qB+m_q^2}{e_qB} K_2 \left( \frac{n}{T} \sqrt{e_q B + m^2_q}\right) - \frac{ne_q
B}{12T} K_0\left( \frac{n}{T} \sqrt{m^2_q + e_q B}\right)\right\}.\label{A5}\ee

One can see in (\ref{A5}), that the sum over  $n$ is well saturated by the
first term with $n=1$ (a typical situation for $\mu=0)$ and for $m_q\to 0$ one
can write \be P_q (B) \approx \frac{ N_c e_q BT}{\pi^2} \bar L \left\{ T + 2T
K_2 \left( \frac{\sqrt{e_q B}}{T} \right) - \frac{e_qB}{12T} K_0 \left(
\frac{\sqrt{e_qB}}{T}\right)\right\}.\label{A6}\ee

For large $e_q B, e_qB\gg T^2$, the first term in the curly brackets dominates
and one returns to Eq. (\ref{47}) (with $\mu=0$).

In the opposite case, $e_qB \ll T^2$ one can write, expanding $K_n (z)$ at
small $z$, \be P_q (B) = \frac{N_c \bar L}{\pi^2} \left\{ 4T^4 + (e_q
B)^2\left[ \frac{1}{6}\ln \left(\frac{2T}{\sqrt{e_qB}}\right)+\frac{3}{16} -\frac{C}{6}\right]
\right\}.\label{A7}\ee Here $C=0.577$ is the Euler constant.

At the point $e_q B=T^2$ one has $K_2(1) =1.625; K_0 (1) =0.421$ and the
coefficient of $\frac{N_c  e_qBT^2\bar L}{\pi^2} = \frac{N_c   T^4 \bar
L}{\pi^2} $ is 4.21 instead of 4 in the limiting form for $e_qB\to 0$, which
implies that the weak field asymptotics $4T^4$ has the 5\% accuracy at $e_q B
=T^2$, whereas the form (\ref{A7}) yields at this point the 0.2\% accuracy.
Therefore to treat the whole intermediate region $e_q B\la T^2$ and $e_qB >T^2$
with few percent accuracy one can use Eq. (\ref{A6}).

For $\mu>0$ one can rewrite $\sum\limits_{n_\bot, \sigma} \chi (\mu)$ in (\ref{25}) in
the same way, as it was  done in (\ref{A2}),

$$ \sum_{n_\bot, \sigma} \chi(\mu) = \int \frac{dp_z}{2\pi} \left\{ \ln \left(1
+ \exp \left(\frac{\bar \mu- \sqrt{p^2_z+ m^2_q}}{T}\right)\right)\right.+$$
\be + 2 \sum^\infty_{n_\bot =0} \ln \left(1+\exp \left( \frac{\bar
\mu-\sqrt{p^2_z + m^2_q +e_q B +2 e_q B (n_\bot +\frac12)}}{T}\right)\right)=
I_1+I_2.\label{A8}\ee

The first term $I_1$ is easily (integrating by parts) transformed to the m.f.
independent term $\phi(\mu)$ in (\ref{52}), \be I_1 = \frac{1}{\pi
T}\phi(\mu).\label{A9}\ee

The second term $I_2$ can be rewritten using (\ref{A3}) as follows
$$ I_2 = \frac{1}{\pi T}\int^\infty_0 \frac{p^2_z dp_z}{e_q B} \int^\infty_0
\frac{d\lambda}{\sqrt{p^2_z+m^2_q+e_qB+\lambda}}\frac{1}{e^{\frac{\sqrt{p^2_z+m^2_q+e_qB+\lambda}-\bar\mu}{T}_{+1}}}-$$
\be - \frac{e_qB}{24 \pi T} \int^\infty_{-\infty}
\frac{dp_z}{\sqrt{p^2_z+m^2_q+e_qB}}\frac{1}{\exp\left(\frac{\sqrt{p^2_z+m^2_q+e_qB}-\bar\mu}{T}\right)+1}\equiv
I'_2+I^{\prime\prime}_2.\label{A10}\ee

One can estimate the large $eB$ asymptotics of $I_2$,$ I_2 \sim \exp
\left(-\frac{\sqrt{e_qB+m^2_q}}{T}\right)$, which supports Eq. (\ref{52}) in
this  limit. In the opposite limit, $eB\to 0$ the first term $I'_2$ in
(\ref{A10}) behaves as \be I'_2 \approx \frac{2}{3\pi T e_qB} \int^\infty_0
\frac{p^4 dp}{\sqrt{p^2+\tilde m^2_q}}\frac{1}{\exp\left(\frac{\sqrt{p^2+\tilde
m^2_q}-\bar \mu}{T}\right)+1},\label{A11}\ee where $\tilde m^2_q = m^2_q +
e_qB$. One can see, that $I'_2(e_qB\to 0)$ ensures in (\ref{25}) the correct
limiting form found in \cite{27}, namely

\be  P_q = \frac{N_cT^4}{3\pi^2}  \left[ \phi_\nu
\left(\frac{\mu-\frac{V_1}{2}}{T}\right) +\phi_\nu
\left(-\frac{\mu+\frac{V_1}{2}}{T}\right)\right], ~~ P= \sum^{n_f}_{q=1} P_q
\label{A12}\ee with \be \phi_\nu (a) = \int^\infty_0 \frac{z^4 dz}{\sqrt{
z^2+\nu^2}} \frac{1}{\exp(\sqrt{z^2 + \nu^2}-a)+1}, \label{A13}\ee and $\nu =
\frac{\sqrt{m^2_q+e_qB}}{T}$. At the same time the term $I^{\prime\prime}_2$
has  the order $O\left(\frac{(e_q B)^2}{T^4}\right)$, as compared to the
leading term (\ref{A12}).

\newpage


\end{document}